# Identification of a monoclinic metallic state in $VO_2$ from a modified first-principles approach

Yongcheng Liang[1], Ping Qin[2], Zhiyong Liang[1], Xun Yuan[1], Yubo Zhang[3]

[1]*College of Science, Donghua University, Shanghai 201620, China*

[2]*College of Engineering Science and Technology, Shanghai Ocean University, Shanghai 201306, China*

[3]*Department of Physics and Engineering Physics, Tulane University, New Orleans 70118, USA*

Metal-insulator transition underlies many remarkable and technologically important phenomena in $VO_2$. Even though its monoclinic structure had before been the reserve of the insulating state, recent experiments have observed an unexpected monoclinic metallic state. Here we use a modified approach combining first-principles calculations with orbital-biased potentials to reproduce the correct stability ordering and electronic structure of different phases of $VO_2$. We identify a ferromagnetic monoclinic metal that is likely to be the experimentally observed mysterious metastable state. Furthermore, our calculations show that an isostructural insulator-metal electronic transition is followed by the lattice distortion from the monoclinic structure to the rutile structure. These results not only explain the experimental observations of the monoclinic metallic state and the decoupled structural and electronic transitions of $VO_2$, but also provide a understanding for the metal-insulator transition in other strongly correlated *d* electron systems.

## 1. Introduction

Vanadium dioxide ($VO_2$) has long been a research focus, not only because it is an archetypical example of metal-insulator transition (MIT),[1-16] but also because the large change of conductivity and optical properties with its MIT brings many potential applications, such as field effect transistors, ultrafast optoelectronic switches and smart windows.[17-21] At about 340 K, the MIT happens in $VO_2$, accompanied by a nearly simultaneous structural distortion from a high-temperature metallic phase with a rutile (R, space group $P4_2/mnm$) structure [see Fig. 1(a)] to a low-temperature insulating phase with a monoclinic (M, space group $P2_1/c$) structure [see Fig. 1(b)]. Nevertheless, the physical origin of this MIT has puzzled scientists for decades. The key issue is whether the M phase is a Peierls insulator driven by the electron-lattice interplay or a Mott insulator driven by the strong electron-electron correlation.[22-29] Recent advances in ultrafast electron diffractions brought much excitement and inspired a new wave of experimental investigations on $VO_2$ with unprecedented accuracy.[3-12] Unexpectedly, a monoclinic metallic state was observed by photoexciting the monoclinic insulating state, indicating an isostructural insulator-metal electronic transition with the decoupling of the subsequent structural transition.[6-8] These observations provide the important evidence to understand the physical mechanism of the MIT at 340 K. However, until now, the unusual monoclinic metallic phase remains elusive.

Great efforts has been made to understand the electronic and structural phase transitions in VO$_2$. According to the simple band model,[23] their electronic properties are well understood. In the R phase, the V atoms are symmetrically equivalent and form one-dimensional chains with uniform bond lengths of 2.85 Å. Each V$^{4+}$ ion has an unpaired valence electron shared by three nearly degenerate $t_{2g}$ orbitals (i.e., $d_{x^2-y^2}$, $d_{xz}$, and $d_{yz}$), and the electron hops near the V site resulting in a metallic state. In the M structure, the V atoms dimerize into the slightly zigzagged chains with alternative lengths of 2.62 and 3.16 Å. The $t_{2g}$ orbitals split into the fully occupied $a_{1g}$ orbitals (i.e., $d_{x^2-y^2}$) and the empty $e_g^\pi$ orbitals (i.e., $d_{xz}$ and $d_{yz}$). As a result, a band gap of about 0.7 eV opens between the $a_{1g}$ and $e_g^\pi$ orbitals [see Fig. 1(c)]. However, multiple active degrees of freedom (lattice, charge, orbital, and spin) are involved to drive the cooperative phase transitions, which challenges the predictability of conventional calculation methods. For example, popular density functional theory (DFT) calculations within local density approximation (LDA) cannot reproduce the M ground state and its insulating nature,[30, 31] Some sophisticated methods (e.g., LDA+$U$, HSE)[32-38] can open a band gap in the M phase, but they give a wrong relative stability ordering of different phases of VO$_2$. Therefore, it is highly desirable to present a valid method to correctly describe the experimental observations and to capture the critical roles of different degrees of freedom in the phase competition of VO$_2$.

In this Letter, we combine first-principles calculations with orbital-dependent potentials to examine the subtle VO$_2$ system. After having reproduced the correct stability ordering and electronic structure of various phases of VO$_2$, we predict a metastable ferromagnetic (FM) metallic M phase that possibly is the experimentally observed intermediate state. Moreover, our results explain the experimental observations that the transition from the M phase to the R phase is initiated by a fast isostructural electronic transition from the nonmagnetic (NM) insulator to the FM metal of the M structure, followed by a slow lattice distortion from the M structure to the R structure.

## 2. The orbital-dependent LDA+$\Delta V$ approach

The conventional LDA method fails to open a band gap for the M phase of VO$_2$. On one hand, it is well known that the LDA method itself underestimates the band gap especially in narrow-gap semiconductors. On the other hand, in VO$_2$, the strong self-interaction error[39] makes the V-3$d$ orbitals very delocalized. Unfortunately, the LDA method usually makes these orbitals too isotropic, and this is also confirmed by our LDA calculations (the occupation number 0.36, 0.32, and 0.32 electrons for the $d_{x^2-y^2}$, $d_{xz}$, and $d_{yz}$ orbitals, respectively). The underestimation of orbitals anisotropy, together with the overestimation of band dispersion, leads to a small overlap between the valence bands (derived from the $a_{1g}$ orbitals) and the conduction bands (derived from the $e_g^\pi$ orbitals).

The LDA+$U$ method,[40] originally proposed for the strongly correlated electrons, is now used to correct the self-interaction error. Although the LDA+$U$ correction opens a gap in the M phase, it produces a wrong relative stability ordering of different magnetostructural phases. Usually, a large Hubbard $U$ value will falsely suppress the hybridization between V-3$d$ and O-2$p$ states. Moreover, the V-3$d$ orbitals of VO$_2$ have an anisotropic feature, and the LDA+$U$ method is insufficient to effectively describe these orbitals differences. In principle, a rigorous description of VO$_2$ should capture the strongly correlated many-body effects and this goes beyond the static one-particle approximation. Much efforts has been made along this line. For example, more sophisticated methods (e.g., GW and QMC)[41, 42] including the many-body effects have been applied to this system. Interestingly, the dynamical mean field theory (DMFT)[43-45] successfully describes the electronic structure of several phases of VO$_2$ and stresses the contributions of dynamical correlations to opening a gap in the M phase. Tomczak *et al.*[46] find that the essential of the full many-body spectrum from DMFT can be reproduced by an orbital-dependent but static one-particle potential (called the LDA+Δ approach).

Our first-principles calculations on VO$_2$ are performed through employing the Vienna *ab initio* simulation package (VASP)[47] and adopting the projector augmented wave (PAW) potentials[48] and the LDA functional[49] for the exchange-correlation energy with a plane-wave basis set. We use the experimental structures of the M and R phases[50, 51] for all calculations. A cutoff energy of 600 eV and 11×11×11 uniform Monkhorst-Pack $k$-grids[52] are taken for the M structure. The tetragonal R structure is transformed into a similar monoclinic supercell and the same cutoff energy and $k$-grids are adopted.

In order to simulate the orbitals anisotropy induced by the self-interaction error, we introduce a small external potential ($\Delta V$) that will bias the selected $e_g^\pi$ orbitals ($d_{xz}$ and $d_{yz}$) to tune their occupations,[53] restoring the correct ordering of the phase stability (including structural, electronic and magnetic phases) that agrees with experiments. This potential adds to the conventional LDA functional, so our calculated functional becomes:

$$E_{\text{LDA}+\Delta V} = E_{\text{LDA}} - \Delta V (n_{xz} + n_{yz}), \quad (1)$$

where $n_{xz}$ and $n_{yz}$ are the occupation numbers of the $d_{xz}$ and $d_{yz}$ orbitals. Accordingly, a new orbital-related term is added to the Kohn-Sham equation:

$$[H_{\text{LDA}} - \Delta V (|d_{xz}\rangle\langle d_{xz}| + |d_{yz}\rangle\langle d_{yz}|)]|\psi_{nk}\rangle = \varepsilon_{nk}|\psi_{nk}\rangle. \quad (2)$$

Obviously, a negative potential will shift the corresponding states upwards and thus punish the $d_{xz}$ and $d_{yz}$ orbital occupations. A positive potential has the opposite effects. This approach (called the LDA+$\Delta V$) has two main advantages. Firstly, the LDA+$\Delta V$ approach only affects the selected $d_{xz}$ and $d_{yz}$ orbitals and has no direct effects on other $d$ orbitals. However, the LDA+$U$ method affects all $d$ orbitals and strongly favor the formation of local magnetic moments. Secondly, in the LDA+$\Delta V$ approach, a small orbital-dependent potential can well correct the anisotropy of different $d$ orbitals and this avoids the false suppression of the $pd$ hybridization. We would like to mention that our method goes beyond an effective low-energy description, although it is phenomenological in nature with a small empirical $\Delta V$ parameter.

## 3. Results and discussion
## 3.1. Relative stability and phase transitions

Figure 2 presents our calculated relative total energies for the R and M structures with different magnetic states with respect to the external potentials on the $e_g^\pi$ orbitals. At $\Delta V = 0$ eV, corresponding to the popular LDA calculation, the FM R phase is predicted to be the most stable phase of VO$_2$. Although the result agrees with the previous first-principles calculations,[30, 31] it fails to reproduce the experimental observation that the NM M phase is the ground state.[22-25] We notice that the relative energy stability of different phases is sensitive to the perturbation potentials on the $e_g^\pi$ orbitals. When introducing the potentials of about the range $\Delta V < -0.5$ eV, our LDA+$\Delta V$ method not only correctly predicts the stability ordering of different magnetostructural phases but also successfully achieves reasonable electronic structures of these phases (see discussion below). These good agreements between our calculations and the experiments manifest that the present method is really effective to deal with the delicate VO$_2$ system.

As can be seen from Fig. 2, the NM M and FM R phases strongly compete with each other. When the potentials are within the range of $\Delta V < -0.5$ eV, the NM M phase is the energetically most favorable phase and becomes the ground state of VO$_2$. With the potentials gradually increasing to $\Delta V \geq -0.5$ eV, the energy of the FM R phase monotonically falls below that of the NM M phase. These results are consistent with the experimental observations that VO$_2$ exhibits a first-order phase transition from the M structure to the R structure at certain conditions (e.g., the temperature of 340 K).[1, 2]

For the M structure, the antiferromagnetic (AFM) state always converges to the NM state, irrespective of the change of the potentials. Interestingly, the FM state is sensitive to the potentials. Our calculations show that the FM state can be viable within $\Delta V > -0.25$ eV, though it also converges to the NM state within $\Delta V \leq -0.25$ eV. We surprisingly find that the FM state becomes more stable in energy than the NM state with the increase of the potentials. This infers that a sudden and rapid variation of the potentials could induce an instantaneous phase transition from the NM state to the FM state while the crystal symmetry completely retains the monoclinic M structure. For the R structure, the FM state is always lower in energy than the NM state, and the AFM state can be obtained only within the range $\Delta V < -0.2$ eV. We note that the energy of the FM R phase is much lower than that of the FM M phase within $\Delta V > -0.25$ eV, thus the metastable FM M phase will eventually transform to the FM R phase.

### 3.2. Electronic mechanisms

We start to analyze the key roles of multiple active degrees of freedom in the phase competition of VO$_2$. At this stage, the following fundamental issues are to be addressed: (1) Why does the NM R phase always exhibit instability? (2) What mechanism stabilizes the NM M phase (the FM R phase) within $\Delta V < -0.5$ eV (within $\Delta V \geq -0.5$ eV)? (3) What drives the isostructural phase transition from the NM M state to the FM M state? The answers to these questions should lie in the fundamental structural and electronic properties of VO$_2$.

Figure 3(a)-3(c) present density of states (DOS) and band structures for the NM R phase at $\Delta V = -0.6$ eV. In the NM R phase, the crystal field effect of the slightly distorted VO$_6$ octahedron approximately splits the V-3$d$ levels into the $t_{2g}$ states and the $e_g$ states. The lowest 24 bands in the range of (-8, -1) eV can be mainly attributed to O-2$p$ states. The 12 bands in the range of (-1, 2) eV are dominated by the $t_{2g}$ states while the 8 bands in the range of (2, 5) eV are derived from the $e_g$ states. In convention LDA calculations, the $d_{x^2-y^2}$ ($d_{xz}$ and $d_{yz}$) orbital occupation in the R phase is underestimated (overestimated), thus the three $t_{2g}$ orbitals have roughly equal occupation. When a small negative potential ($\Delta V = -0.6$ eV) is introduced to redistribute the $t_{2g}$ states, the most significant difference is that the $d_{x^2-y^2}$ orbital is a bit separated from the higher $d_{xz}$ and $d_{yz}$ orbitals, and this significant correction was also achieved by the DMFT method.[45] We can clearly see that the Fermi level crosses through the $t_{2g}$ states and lies nearly at the peak of the DOS curve. It is this high DOS at the Fermi level that results in the instability of the NM R phase of VO$_2$. Usually, there are two fundamental modes for restoring the stability of a structure with a high DOS at the Fermi level: one is by structural distortion (the Peierls-like mechanism) and another is by developing magnetism (the Stoner-like mechanism).[54]

At $\Delta V < -0.5$ eV, the NM M phase becomes the ground state and the Peierls-like mechanism dominates. When the tetragonal R structure is distorted into the monoclinic M structure, the V atoms dimerize to form slightly zigzagged chains [see Figs. 1(a) and 1(b)]. Because of the dimerization, the $a_{1g}$ states are split into fully occupied bonding states and unoccupied antibonding states as shown in Fig. 3(d)-3(f). Meanwhile, the $e_g^\pi$ states become fully empty. As a result, a band gap successfully opens. Hence, the structural distortion breaks the degeneracy of the $t_{2g}$ orbitals, which plays an important role in the stability of the NM M phase. Although the dimerization has been considered as a hallmark of the Peierls transition, we would like to emphasize that the NM M insulator is different from other pure Peierls insulators (e.g. MnB$_4$, K$_2$Cr$_8$O$_{16}$)[54, 55] since our calculations include the small external potentials. More recently, a DFT+DMFT perspective points to the dynamical electron correlations acting in collaboration with the lattice distortion as being responsible for opening the gap in the NM M state.[43] If the punishing effect of our introduced potentials is imagined as the DFT+DMFT self energy in their work, our results support their view.

When the potentials gradually increase to $\Delta V \geq -0.5$ volt, the Stoner-like mechanism is more effective in stabilizing the FM R phase. The electronic instability of the NM R state can be removed by breaking the spin symmetry to develop magnetism. In particular, the developing of the FM ordering reduces the kinetic energy of V-3$d$ electrons to stabilize the system. The energy gain in the FM R phase is also captured by our LDA+$\Delta V$ calculations: the FM R state is always more stable than the AFM R and the NM R states (see Fig. 2). Although this FM ordering leads to a large depletion of DOS at the Fermi level, the FM R phase does not develop a gap and remains metallic [see Figs. 4(a) and 4(b)]. Hence, the crossover between the two phases (NM M and FM R) is a manifestation of the strong competition of the Peierls-like mechanism and the Stoner-like mechanism.

As the potentials increase to $\Delta V > -0.25$ eV, the NM M state may transform to the FM M state while the system remains monoclinic. It thus is of great interest to unravel this competition of two magnetic orderings with the same M structure. Figure 3(g)-3(i) display the DOS and band structures for the NM M phase at $\Delta V = 0.3$ eV. The increase of the potentials makes the redistribution of the electronic occupation of the $t_{2g}$ orbitals. In particular, the Fermi level of the NM M phase lies nearly at the peak position of the $a_{1g}$ bonding states. This high DOS at the Fermi level strongly drives the NM state towards the FM instability. Figure 4(c) and 4(d) show the DOS and band structures, respectively, of the $t_{2g}$ states of the FM M phase at $\Delta V = 0.3$ eV. Compared with the NM M phase, the spin polarization in the FM M phase causes a large spin splitting of the $t_{2g}$ states. A prominent feature is the separation of the $a_{1g}$ bonding states into the majority and minority spins. Although the majority spin $a_{1g}$ bonding states still lie in below the Fermi level, the corresponding minority spin states are lifted above the Fermi level. At the same time, the majority spin $e_g^\pi$ states become partially occupied. This redistribution of electronic occupation of the $t_{2g}$ orbitals leads to the FM M phase to become metallic. Considering that the M structure beforehand had universally been the reserve of the insulating state, it is rather surprising that the FM M phase exhibits metallicity. Therefore, the isostructural crossover from the NM M state to the FM M state is an insulator-metal electronic transition driven by the Stoner-like mechanism.

As can be seen from Fig. 2, the energy of the FM R phase is much lower than that of the FM M phase. Thus this insulator-metal crossover from the NM M state to the FM M state is an intermediate electronic transition, and the metastable FM M state eventually transforms to the energetically more favorable FM R state. As discussed above, the redistribution of the $t_{2g}$ orbital occupation is responsible for the electronic transition. Note that at this stage, however, the lattice remains monoclinic. When the external potentials gradually weaken the V-V pairs, the electron-lattice coupling and thermalization eventually eliminate the zigzag arrangement of the V atoms and the structural phase transition takes place from the FM M state to the FM R state. Therefore, the MIT of $VO_2$ can be decoupled into the electronic transition from the NM M state to the FM M state and the structural transition from the FM M state to the FM R state, which are separated by the FM M phase.

Several experiments[6-8] have recently discovered a monoclinic metallic phase during the application of femtosecond laser pulses on $VO_2$ samples. For example, Morrison et al.[6] decoupled the electronic and lattice degrees of freedom using the ultrafast diffraction experiment. By photoexciting the monoclinic insulating M phase, they observe a new monoclinic metallic state which is caused the 'heated' electrons. The 'heated' electrons have a high intensity to hop between the $t_{2g}$ orbitals, and make these orbitals more homogeneous. In our LDA+$\Delta V$ approach, a positive potential on the $e_g^\pi$ orbitals has an effect to diminish the differences between the $a_{1g}$ and $e_g^\pi$ orbitals. Figure 2 shows an isostructural phase transition from the NM M state to the FM M state when the splitting between the $a_{1g}$ and $e_g^\pi$ orbitals is suppressed with the increase of the potentials. The reduction of this splitting is driven by the Stoner-like mechanism, which is related to the fast electronic dynamics. With the weakening of the V-V pairs, the Peierls-like mechanism starts to dominate the transition, which is related to the slow ionic dynamics. Eventually, the V-V pairs are destroyed, and the transition from the M structure to the R structure is realized. These results highly accord with the experimental findings that the isostructural insulator-metal electronic transition is followed by the structural transition. Our calculations therefore provide a satisfactory explanation for the experimental observations of the monoclinic metallic state and the decoupled electronic and structural components of photoinduced phase transitions of $VO_2$.

## 4. Conclusions

In summary, we combine first-principles calculations with orbital-dependent perturbations to evaluate the phase stability and electronic structures of the delicate $VO_2$ system. By introducing the suitable potentials to adjust the occupation of the V-3$d$ orbital, the experimental ground-state NM M phase can be correctly obtained and its insulating nature can also be reproduced. More importantly, we find that monoclinic $VO_2$ can undergo an intermediate isostructural transition from the insulating NM M phase to the metallic FM M phase, which highly accords with the experimental discovery of the metallic state with the monoclinic symmetry. We also reveal that such an isostructural insulator-metal transition is a result of the electronic redistribution of the V-3$d$ derived $t_{2g}$ orbital occupation.

**Acknowledgements**

We much thank Prof. Peihong Zhang and Prof. Haijun Zhang for their valuable discussion. We also acknowledge financial support from the National Natural Science Foundation of China (No. 51671126).

**References**

1. J. H. Park, J. M. Coy, T. S. Kasirga, C. Huang, Z. Fei, S. Hunter, and D. H. Cobden, *Nature* **500** (2013) 431.

2. J. D. Budai, J. Hong, M. E. Manley, E. D. Specht, C. W. Li, J. Z. Tischler, D. L. Abernathy, A. H. Said, B. M. Leu, L. A. Boatner, R. J. McQueeney, and O. Delaire, *Nature 515* (2014) 535.

3. B. T. O'Callahan, A. C. Jones, J. H. Park, D. H. Cobden, J. M. Atkin, M. B. Raschke, and B. Markus, *Nat. Commun.* **6** (2015) 6849.

4. M. M. Qazilbash, M. Brehm, B. –G. Chae, P. –C. Ho, G. O. Andreev, B. –J. Kim, S. J. Yun, A. V. Balatsky, M. B. Maple, F. Keilmann, H. –T. Kim, and D. N. Basov, *Science* **318** (2007) 1750.

5. P. Baum, D. –S. Yang, and A. H. Zewail, *Science* **318** (2007) 788.

6. V. R. Morrison, R. P. Chatelain, K. L. Tiwari, A. Hendaoui, A. Bruhács, M. Chaker, and B. J. Siwick, *Science* **346** (2014) 445.

7. D. Wegkamp, M. Herzog, L. Xian, M. Gatti, P. Cudazzo, C. L. McGahan, R. E. Marvel, R. F. Haglund, Jr., A. Rubio, M. Wolf, and J. Stähler, *Phys. Rev. Lett.* **113** (2014) 216401.

8. J. Laverock, S. Kittiwatanakul, A. A. Zakharov, Y. R. Niu, B. Chen, S. A. Wolf, J.W. Lu, and K. E. Smith, *Phys. Rev. Lett. 113* (2014) 216402.

9. M. K. Liu, M. Wagner, E. Abreu, S. Kittiwatanakul, A. McLeod, Z. Fei, M. Goldflam, S. Dai, M. M. Fogler, J. Lu, S. A. Wolf, R. D. Averitt, and D. N. Basov, *Phys. Rev. Lett.* **111** (2013) 096602.

10. E. Arcangeletti, L. Baldassarre, D. Di Castro, S. Lupi, L. Malavasi, C. Marini, A. Perucchi, and P. Postorino, *Phys. Rev. Lett.* **98** (2007) 196406.

11. W.-P. Hsieh, M. Trigo, D. A. Reis, G. A. Artioli, L. Malavasi, and W. L. Mao, *Appl. Phys. Lett.* **104** (2014) 021917.

12. A. Cavalleri, Cs. Tóth, C. W. Siders, J. A. Squier, F. Ráksi, P. Forget, and J. C. Kieffer, *Phys. Rev. Lett.* **87** (2001) 237401.

13. H .-T. Kim, Y. W. Lee, B. -J. Kim, B.- G. Chae, S. J. Yun, K.- Y. Kang, K. -J. Han, K. -J. Yee, and Y. -S. Lim, *Phys. Rev. Lett.* **97** (2006) 266401.

14. Z. Tao, Tzong-Ru T. Han, S. D. Mahanti, P. M. Duxbury, F. Yuan, C. -Y. Ruan, K. Wang, and J. Wu, *Phys. Rev. Lett.* **109** (2012) 166406.

15. M. Mitrano, B. Maroni, C. Marini, M. Hanfland, B. Joseph, P. Postorino, and L. Malavasi, *Phys. Rev. B* **85** (2012) 184108.

16. C. Marini1, S. Pascarelli1, O. Mathon1, B. Joseph, L. Malavasi, and P. Postorino, *Europhys. Lett.* **102** (2013) 66004.

17. B. Hu, Y. Ding, W. Chen, D. Kulkarni, Y. Shen, V. V. Tsukruk, and Z. L. Wang, *Adv. Mater.* **22** (2010) 5134.

18. M. Nakano1, K. Shibuya1, D. Okuyama, T. Hatano, S. Ono, M. Kawasaki, Y. Iwasa, and Y. Tokura, *Nature* **487** (2012) 459.

19. N. B. Aetukuri, A. X. Gray, M. Drouard, M. Cossale, L. Gao, A. H. Reid, R. Kukreja, H. Ohldag, C. A. Jenkins, E. Arenholz, K. P. Roche, H. A. Dürr, M. G. Samant, and S. S. P. Parkin, *Nature Phys.* **9** (2013) 661.

20. J. Jeong, N. Aetukuri, T. Graf, T. D. Schladt, M. G. Samant, and S. S. P. Parkin, *Science* **339** (2013) 1402.

21. K. Appavoo, B. Wang, N. F. Brady, M. Seo, J. Nag, R. P. Prasankumar, D. J. Hilton, S. T. Pantelides, and R. F. Haglund, *Nano Lett.* **14** (2014) 1127.

22. F. J. Morin, *Phys. Rev. Lett.* **3** (1959) 34.

23. J. B. Goodenough, *J. Solid. State Chem.* **3** (1971) 490.

24. A. Zylbersztejn and N. F. Mott, *Phys. Rev. B* **11** (1975) 4383.

25. M. W. Haverkort, Z. Hu, A. Tanaka, W. Reichelt, S. V. Streltsov, M. A. Korotin, V. I. Anisimov, H. H. Hsieh, H. -J. Lin, C. T. Chen, D. I. Khomskii, and L. H. Tjeng, *Phys. Rev. Lett.* **95** (2005) 196404.

26. T. Yao, X. Zhang, Z. Sun, S. Liu, Y. Huang, Y. Xie, C. Wu, X. Yuan, W. Zhang, Z. Wu, G. Pan, F. Hu, L. Wu, Q. Liu, and S. Wei, *Phys. Rev. Lett.* **105** (2010) 226405.

27. T. L. Cocker, L. V. Titova, S. Fourmaux, G. Holloway, H. -C. Bandulet, D. Brassard, J. -C. Kieffer, M. A. El Khakani, and F. A. Hegmann, *Phys. Rev. B* **85** (2012) 155120.

28. M. van Veenendaal, *Phys. Rev. B* **87** (2013) 235118.

29. X. Yuan, W. Zhang, and P. Zhang, *Phys. Rev. B* **88** (2013) 035119.

30. R. M. Wentzcovitch, W. W. Schulz, and P. B. Allen, *Phys. Rev. Lett.* **72** (1994) 3389.

31. V. Eyert, *Ann. Phys.* **11** (2002) 650.

32. A. Liebsch, H. Ishida, and G. Bihlmayer, *Phys. Rev. B* **71** (2005) 085109.

33. M. S. Laad, L. Craco, and E. Muller-Hartmann, *Phys. Rev. B* **73** (2006) 195120.

34. J. M.Tomczak and S. Biermann, *J. Phys.: Condens. Matter* **19** (2007) 365206.

35. R. Sakuma, T. Miyake, and F. Aryasetiawan, *Phys. Rev. B* **78** (2008) 075106.

36. V. Eyert, *Phys. Rev. Lett.* **107** (2011) 016401.

37. F. Iori, M. Gatti, and A. Rubio, *Phys. Rev. B* **85** (2012)115129.

38. R. Grau-Crespo, H. Wang, and U. Schwingenschlogl, *Phys. Rev. B* **86** (2012) 081101.

39. J. P.Perdew and A. Zunger, *Phys. Rev. B* **23** (1981) 5048.

40. M. Cococcioni and S. Gironcoli, *Phys. Rev. B* **71** (2005) 035105.

41. M. Gatti, F. Bruneval, V. Olevano, and L. Reining, *Phys. Rev. Lett.* **99** (2007) 266402.

42. H. H. Zheng and L. K. Wagner, *Phys. Re. Lett.* **114** (2015) 176401.

43. W. H. Brito, M. C. O. Aguiar, K. Haule, and G. Kotliar, *Phys. Rev. Lett.* **117** (2016) 056402.

44. C. Weber, D. D. O'Regan, N. D. M. Hine, M. C. Payne, G. Kotliar, and P. B. Littlewood, *Phys. Rev. Lett.* **108** (2012) 256402.

45. S. Biermann, A. Poteryaev, A. I. Lichtenstein and A. Georges, *Phys. Rev. Lett.* **94** (2005) 026404.

46. J. M. Tomczak, F. Aryasetiawan, and S. Biermann, *Phys. Rev. B* **78** (2008) 115103.

47. G. Kresse and J. Furthmuller, *Phys. Rev. B* **54** (1996) 11169.

48. G. Kresse and D. Joubert, *Phys. Rev. B* **59** (1999) 1758.

49. D. M. Ceperley and B. J. Alder, *Phys. Rev. Lett.* **45** (1980) 566.

50. M. Marezio, D. B. McWhan, J. P. Remeika, and P. D. Dernier, *Phys. Rev. B* **5** (1972) 2541.

51. D. B. McWhan, M. Marezio, J. P. Remeika, and P. D. Dernier, *Phys. Rev. B* **10** (1974) 490.

52. H. J. Monkhorst and J. D. Pack, *Phys. Rev. B* **13** (1976) 5188.

53. X. Yuan, Y. Zhang, T. A. Abtew, P. Zhang, and W. Zhang, *Phys. Rev. B* **86** (2012) 235103.

54. Y. Liang, X. Yuan, Y. Gao, W. Zhang, and P. Zhang, *Phys. Rev. Lett.* **113** (2014) 176401.

55. T. Toriyama, A. Nakao, Y. Yamaki, H. Nakao, Y. Murakami, K. Hasegawa, M. Isobe, Y. Ueda, A. V. Ushakov, D. I. Khomskii, S. V. Streltsov, T. Konishi, and Y. Ohta, Phys. Rev. Lett. **107**, 266402 (2011).

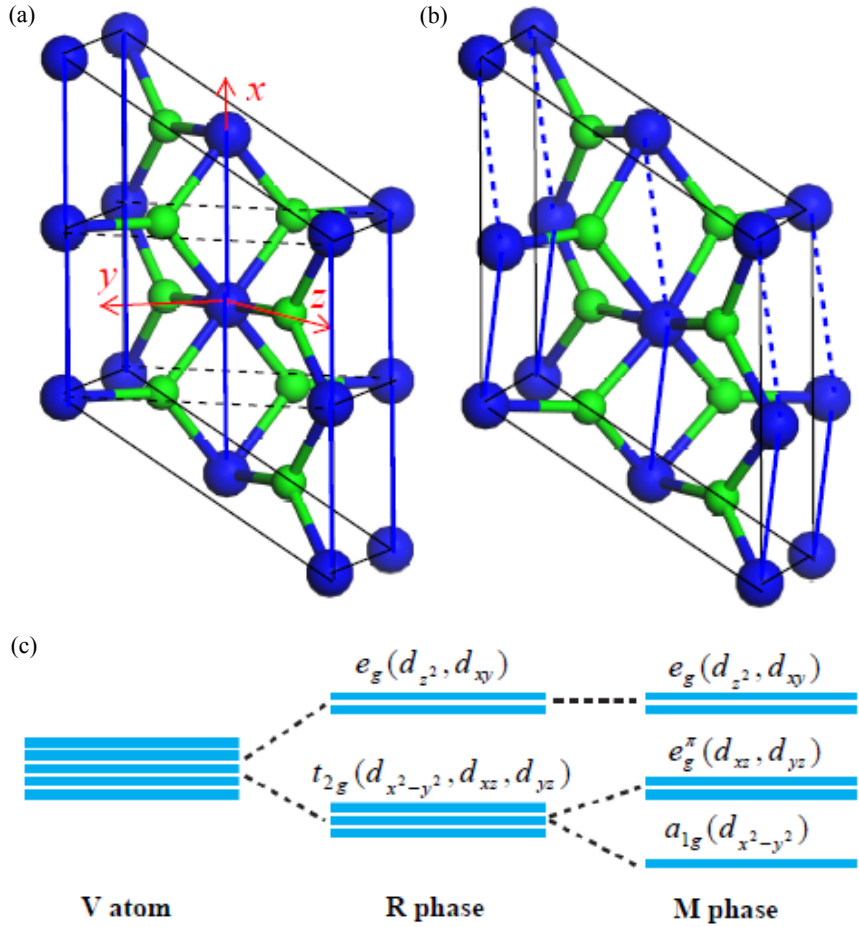

**Fig. 1.** (Color online) Crystal structures of $VO_2$ and the splitting of the V-3$d$ orbitals. (a) the tetragonal R phase, (b) the monoclinic M phase, and (c) the splitting of the V-3$d$ orbitals. The blue (large) and green (small) spheres represent V and O atoms, respectively. The low-symmetry M structure can be derived from the high-symmetry R structure through lattice distortions of the V atoms. In the R structure, the V atoms form one-dimensional chains with uniform V-V distances of 2.85 Å (shown by blue solid lines) along the $x$-axis. In the M structure, the V atoms are dimerized into slightly zigzagged chains with alternating V-V distances of 2.62 Å and 3.16 Å (shown by alternating blue solid and dashed lines).

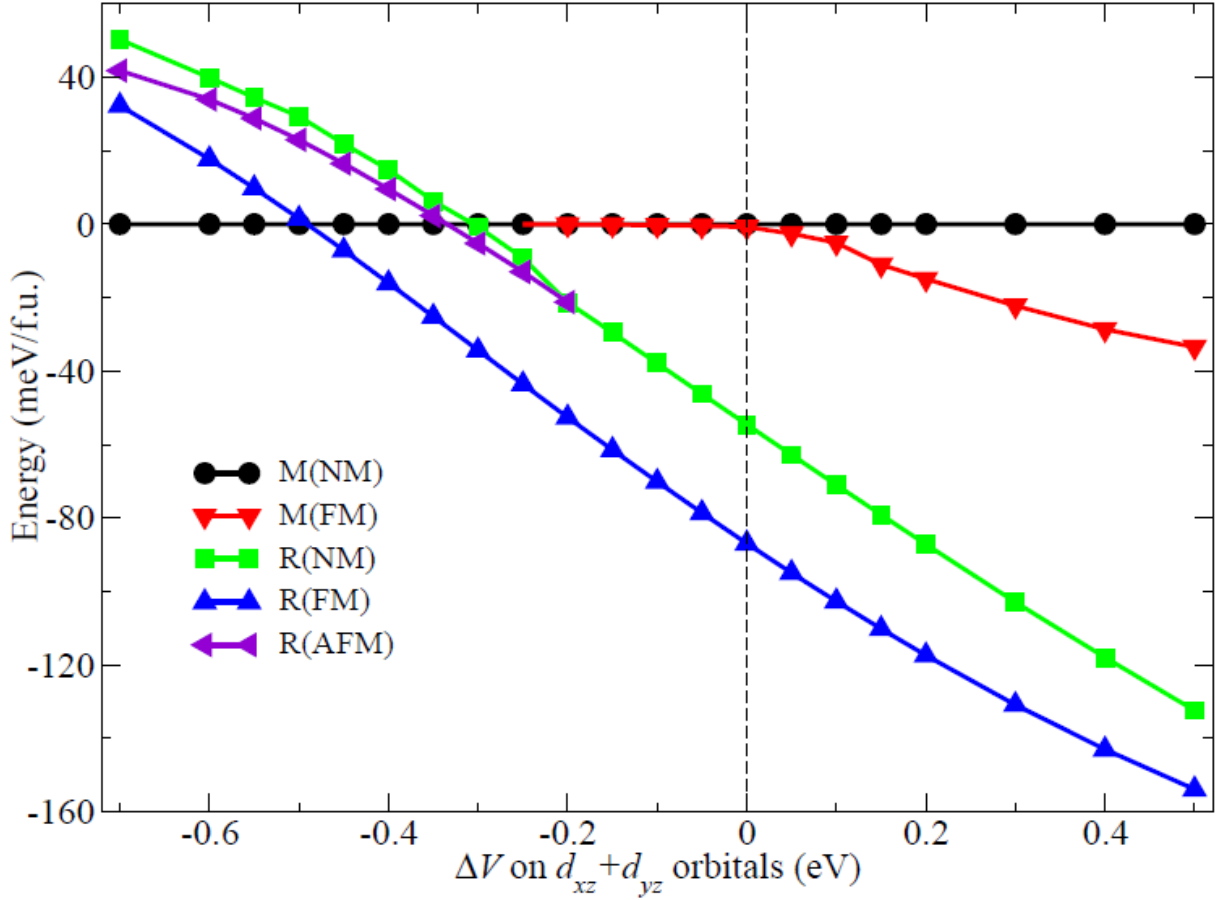

**Fig. 2.** (Color online) The phase stability ordering of VO$_2$. Relative energetic stability of the R and M structures with different magnetic states with respect to the potential ($\Delta V$) on the V-3$d_{xz}$ and V-3$d_{yz}$ orbitals. The energies of the NM M phase are set as the reference energy (i.e., set to zero). All energies are rescaled for one VO$_2$ formula unit (f.u.).

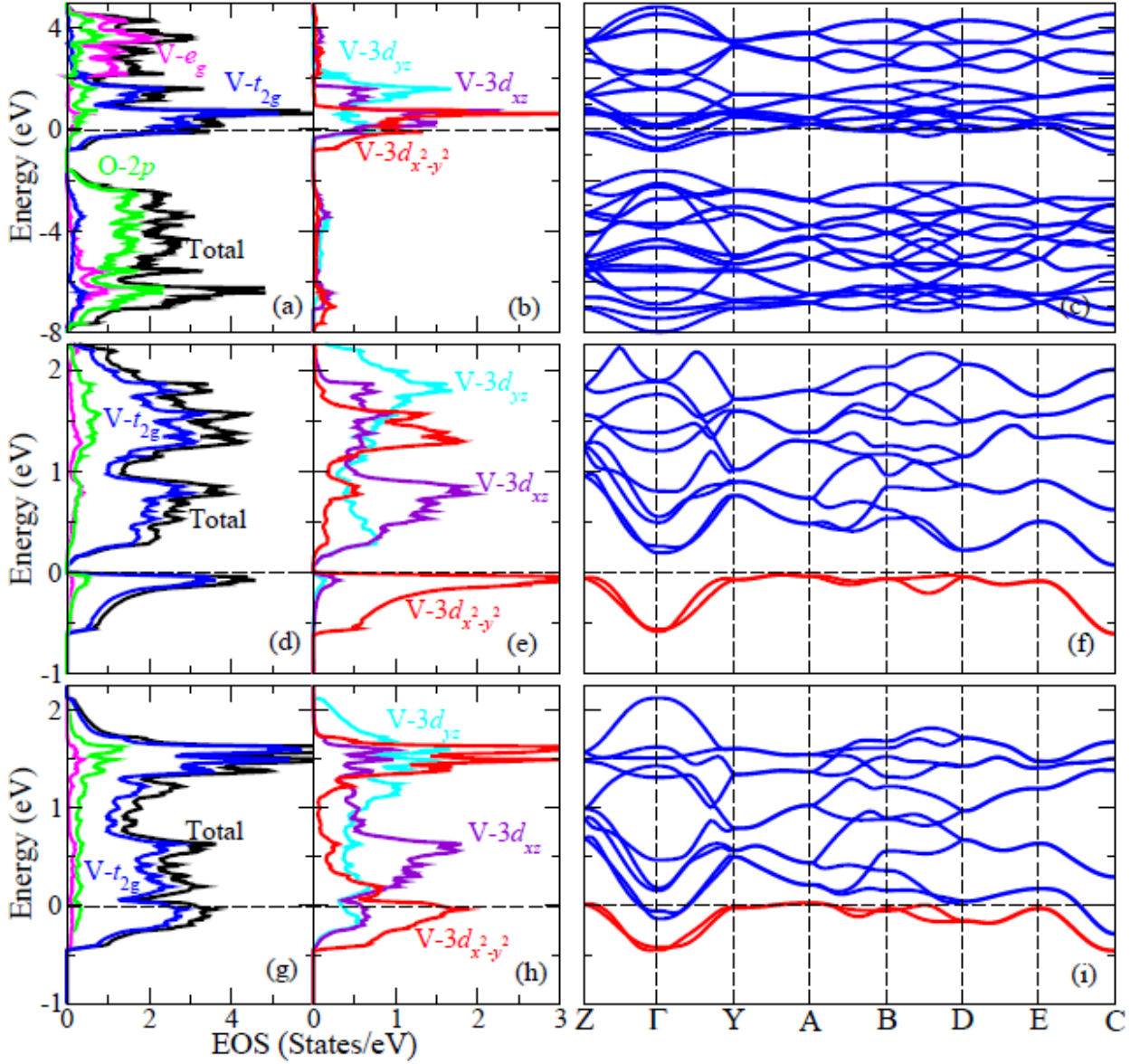

**Fig. 3.** (Color online) Electronic structures of different phases of VO$_2$. Total DOS, projected DOS of V-3$d$ and O-2$p$ orbitals (left); projected DOS of $t_{2g}$ orbitals (middle); and band structures (right) of the NM R phase at $\Delta V = -0.6$ eV (top panels), the NM M phase at $\Delta V = -0.6$ eV (middle panels), and the NM M phase at $\Delta V = 0.3$ eV (bottom panels) with the perturbation method. The Fermi levels are set at 0 eV and shown as horizontal dashed lines.

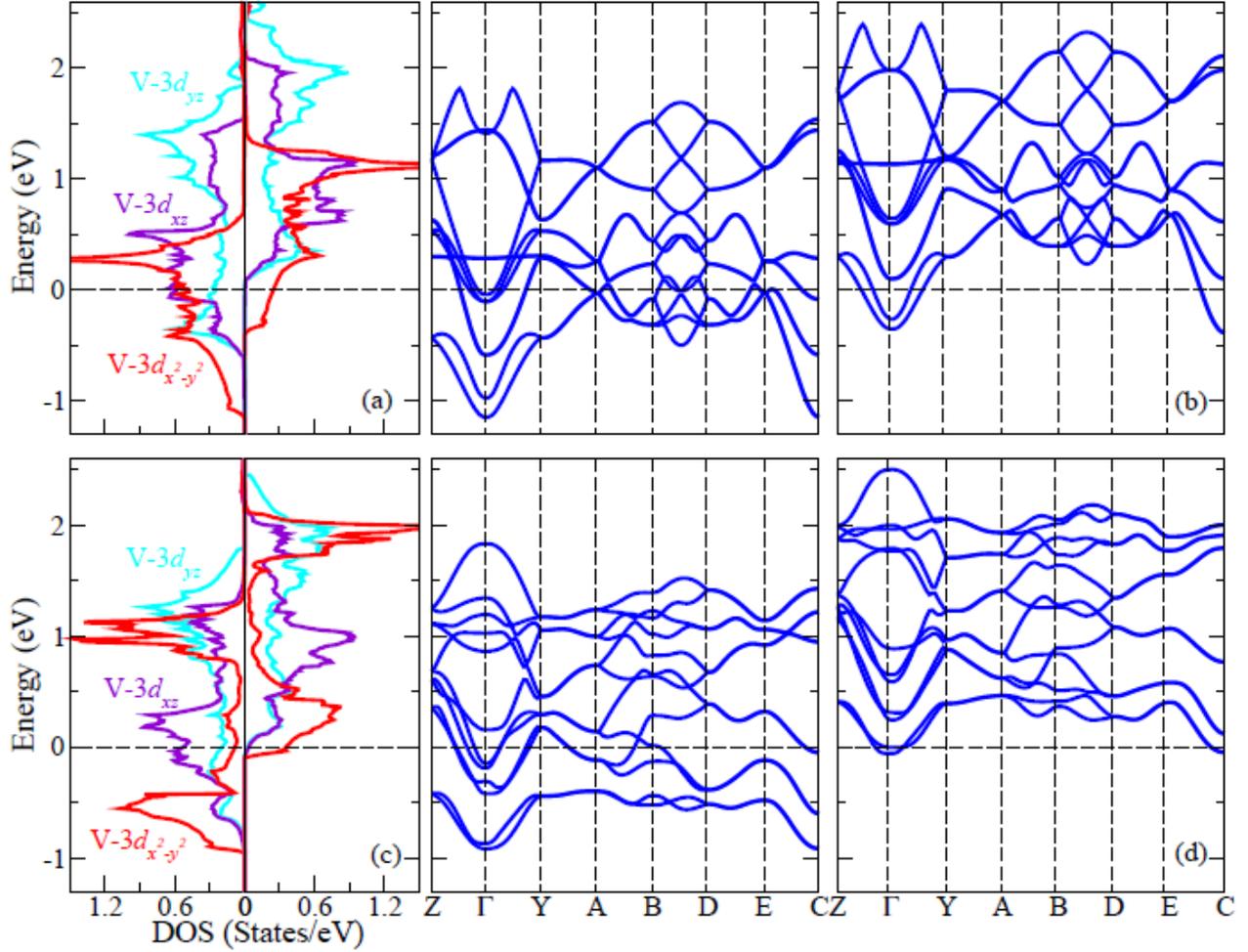

**Fig. 4.** (Color online) Electronic structures of different phases of $VO_2$. Projected DOS and band structures of the FM R phase at $\Delta V = -0.6$ eV (top panels) and the FM M phase (bottom panels) at $\Delta V = 0.3$ eV with the perturbation method. Their left and right panels represent the majority and minority spins, respectively. The Fermi levels are set at 0 eV and shown as horizontal dashed lines.